\documentstyle[preprint,aps,tighten,epsf,floats]{revtex}
\def\Jn#1#2#3#4{{#1} {\bf #2}, #3 (#4)}
\def\PRC{Phys. Rev. C}

\def\PR{Phys. Rev.}
\def\AP{Ann. Phys. (N.Y.)}
\def\NC{Nuovo Cimento}
\def\NCS{Nuovo Cimento Suppl.}
\def\NP{Nucl. Phys.}

\def\RPP{Rep. Prog. Phys.}
\def\IB{{\em ibid.}}
\newcommand{\nn}{\nonumber\\}
\newcommand{\be}{\begin{equation}}
\newcommand{\ee}{\end{equation}}
\newcommand{\bea}{\begin{eqnarray}}
\newcommand{\eea}{\end{eqnarray}}
\newcommand{\bi}{\bibitem}
\newcommand{\A}{{\alpha}}
\newcommand{\B}{{\beta}}

\newcommand{\eqn}[1]{\label{#1}}
\newcommand{\eq}[1]{Eq.~(\ref{#1})}
\newcommand{\eqs}[1]{Eqs.~(\ref{#1})}
\newcommand{\fign}[1]{\label{#1}}
\newcommand{\fig}[1]{Fig.~\ref{#1}}
\newcommand{\figs}[1]{Figs.~\ref{#1}}
\newcommand{\NN}{$N\!N$}
\newcommand{\NNN}{$N\!N\!N$}
\newcommand{\NNNN}{$N\!N\!\rightarrow\!N\!N$}
\newcommand{\piN}{$\pi N$}
\newcommand{\piNN}{$\pi N\!N$}

\newcommand{\NNpid}{$N\!N\!\rightarrow\!\pi d$}
\newcommand{\pidpid}{$\pi d\!\rightarrow\!\pi d$}
\newcommand{\pidNN}{$\pi d\!\rightarrow\!N\!N$}
\newcommand{\pidpiNN}{$\pi d\!\rightarrow\pi\!N\!N$}

\newcommand{\VOPE}{V_{N\!N}^{\mbox{\protect\scriptsize OPE}}}
\newcommand{\ND}{_{\mbox{\protect\scriptsize ND}}}
\newcommand{\F}{{\cal F}}
\newcommand{\M}{{\cal M}}
\newcommand{\V}{{\cal V}}
\newcommand{\G}{{\cal G}}
\newcommand{\T}{{\cal T}}
\newcommand{\tX}{\tilde{X}}

\newcommand{\bphi}{\bar{\phi}}
\newcommand{\f}{\bar{f}}
\newcommand{\bF}{\bar{F}}
\newcommand{\bB}{\bar{B}}
\newcommand{\GpiN}{D_{\pi N}}
\newcommand{\GNN}{D_0}
\newcommand{\VNN}{V_{N\!N}}
\newcommand{\TNN}{T_{N\!N}}
\newcommand{\bT}{{\bar{T}}}
\newcommand{\Ah}{\frac{A}{2}}
\newcommand{\piNNpiNN}{$\pi N\! N\!\rightarrow\!\pi N\!N$}
\newcommand{\NNtopiNN}{$N\!N\!\rightarrow\!\pi N\!N$}

\begin{document}
\draft
\title{Unified relativistic description of \mbox{\boldmath{$\pi N\!N$}}
and \mbox{\boldmath{$\gamma\pi N\!N$}}}


\author{A. N. Kvinikhidze\footnote{On leave from The Mathematical Institute of
Georgian Academy of Sciences, Tbilisi, Georgia.} and B. Blankleider}
\address{Department of Physics, The Flinders University of South Australia,
Bedford Park, SA 5042, Australia}

\maketitle
\begin{abstract}
We present a unified description of the relativistic \piNN\ and $\gamma \pi
N\!N$ systems where the strong interactions are described non-perturbatively by
four-dimensional integral equations. A feature of our approach is that the
photon is coupled in all possible ways to the strong interaction contributions.
Thus the hadronic processes \NNNN, \NNpid, \pidpid, etc.\ and corresponding
electromagnetic processes $N\!N\rightarrow \gamma N\!N$, $\gamma d\rightarrow
N\!N$, $\gamma d\rightarrow \pi d$, $ed\rightarrow e d$, $ed\rightarrow e'\pi
d$, etc., are described simultaneously within the one model of strong
interactions.  Our formulation obeys two and three-body unitarity, and as
photons are coupled everywhere in the strong interaction model, gauge invariance
is implemented in the way prescribed by quantum field theory. Our formulation is
also free from the overcounting and undercounting problems plaguing
four-dimensional descriptions of $\pi N\!N$-like systems. The unified
description is achieved through the use of the recently introduced gauging of
equations method.
\end{abstract}


\section{Introduction}

Recently we have introduced a method for incorporating an external
electromagnetic field into any model of hadrons whose strong interactions are
described through the solution of integral equations \cite{gnnn4d,g4d}. The
method involves the gauging of the integral equations themselves, and results in
electromagnetic amplitudes where an external photon is coupled to every part of
every strong interaction diagram in the model. Current conservation is therefore
implemented in the theoretically correct fashion, i.e.\ as prescribed by quantum
field theory. Initially we applied our gauging procedure to the relativistic
three-nucleon problem whose strong interactions are described by standard
four-dimensional three-body integral equations \cite{gnnn4d,g4d}. More recently
we used the same method to gauge the three-dimensional spectator equation for a
system of three-nucleons \cite{g3d,gnnn3d}.

Here we apply our gauging procedure to the more complicated case of the
relativistic \piNN\ system whose four-dimensional integral equations have only
recently been derived \cite{4d,PA4d}.\footnote{A summary of the present work was
previously reported in a conference proceeding \cite{korea}} These equations
obey both 2-body (\NN) and three-body (\piNN) unitarity and simultaneously
describe all the strong interaction processes of the \piNN\ system, including
the reactions \NNNN, \NNpid, \NNtopiNN, \pidpid\ and \pidpiNN. After gauging
these \piNN\ equations we obtain gauge invariant expressions for all possible
electromagnetic processes of the \piNN\ system, e.g. pion photoproduction
$\gamma d\rightarrow \pi d$ and $\gamma d\rightarrow \pi N\! N$, pion
electroproduction $e d\rightarrow e' \pi d$ and $e d\rightarrow e' \pi N\!N$,
and because pion absorption is taken into account explicitly, we also obtain
gauge invariant expressions for processes like deuteron photodisintegration
$\gamma d\rightarrow N N$ and Bremsstrahlung $NN\rightarrow \gamma NN$ that are
valid even at energies above pion production threshold. Included in the
electromagnetic processes described by our model is the especially interesting
case of elastic electron-deuteron scattering. Here the deuteron is described
within our model as a bound state of the \piNN\ system; thus, after gauging, our
model provides a rich description of the electromagnetic form factors of the
deuteron with all possible meson exchange currents taken into account.
\begin{figure}[t]
\hspace*{2cm} \epsfxsize=13.5cm\epsffile{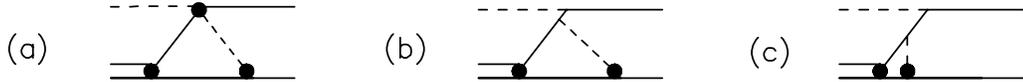}
\vspace{4mm}

\caption{\fign{t} Example of overcounting in \NNpid. (a) The \NNpid\ 
Feynman diagram where dark circles represent all possible contributions.
(b) One of the contributions included in (a). (c) Another way of drawing
diagram (b) showing that this term corresponds to an overdressing of the
deuteron vertex.}
\end{figure} 
The four-dimensional integral equations for the \piNN\ system are a good deal
more complicated than the corresponding equations describing a strictly
three-body system like that of three nucleons. Firstly the \piNN\ equations take
into account pion absorption so there is coupling to the two-body \NN\ channel.
A consequence of this in the four-dimensional approach is that certain
three-body forces must be retained both to avoid the undercounting of diagrams
and to ensure that the equations satisfy two- and three-body unitarity
\cite{4d}. Also, unlike in the \NNN\ case, not all three-body reducible diagrams
of the \piNN\ system have unique three-body cuts. This leads to an overcounting
of diagrams problem in the formulation of four-dimensional \piNN\ equations that
has only recently been solved \cite{4d}. One example of the type of overcounting
encountered is given in \fig{t}.  The newly derived \piNN\ equations have the
feature that their kernels contain explicit subtraction terms which eliminate
all such overcounting.

The presence of pion absorption and the overcounting problem in the \piNN\
system would normally make the task of formulating a description of the
$\gamma$\piNN\ system an especially difficult one. To put this difficulty into
perspective, we recall our experience with gauging the four-dimensional
three-nucleon system \cite{gnnn4d,g4d}. In this case the strong interaction
integral equations have exactly the same form as the three-body equations of
quantum mechanics (QM) (except of course that they are four-dimensional rather
than three-dimensional). Then after applying our gauging procedure we obtained
an electromagnetic current that is again just of the same form as the
three-nucleon current in QM; namely, a simple sum of one and two-body
currents. However, in contrast to the QM case, the two-body current obtained in
the four-dimensional formulation is modified by a subtraction term whose
presence is necessary to stop the overcounting of diagrams. Thus the gauging of
equations method led us to a simple prescription for the electromagnetic
current of three nucleons. The situation for \piNN\ is quite different as the
strong interaction equations for this system cannot be cast into a QM form. Thus
there is no corresponding simple prescription for the \piNN\ electromagnetic
current and the gauging procedure itself becomes effectively the only way to
specify this current. On the other hand, our gauging procedure is extremely
simple, and by gauging the \piNN\ equations with subtraction terms included, one
easily constructs equations for the $\gamma$\piNN\ system without encountering
any further difficulties and with all overcounting problems being taken care of
automatically. In this way we obtain a unified description of the \piNN\ and
$\gamma$\piNN\ systems.

\section{Four-dimensional \mbox{\boldmath{$\pi N\!N$}} equations}

The first attempts to formulate few-body equations using relativistic quantum
field theory were made already in the early 1960's
\cite{Taylor,Tucciarone,Broido}. Both such general formulations and ones more
specific to the \piNN\ system have been pursued until the present time
\cite{Avishai4d,AB4d,Haberzettl}. Yet all these attempts have had theoretical
inconsistencies, including the undercounting and overcounting problems discussed
above. We have recently overcome these problems and derived new consistent
\piNN\ equations by using a method where in cases like that of \figs{t}(b) and
(c) where the rightmost \piNN\ cut is not unique, one of the rightmost \piNN\
vertices is ``pulled out'' further to the right, in this way defining a unique
\piNN\ cut \cite{4d}. The same \piNN\ equations were later derived in Ref.\
\cite{PA4d} where a method based on Taylor's classification of diagrams
\cite{Taylor,PA} was used. In this section we would simply like to restate these
equations but in a form that is particularly convenient for gauging. In this
section we would simply like to restate these equations but in a form that is
particularly convenient for gauging.

\subsection{Distinguishable nucleon case}

Initially we would like to consider the case where the two nucleons are treated
as distinguishable particles. Not only does this avoid taking into account the
symmetry factors and related complications due to the identity of the two
nucleons, but it is also of practical interest in itself, as for example in
modelling the $\pi np$ system without using isospin symmetry.

We follow the usual convention and refer to the two nucleons as particles $1$
and $2$, and the pion as particle $3$. We also use $\lambda=1$ or $2$ to label
the channel where nucleon $\lambda$ and the pion form a two-particle subsystem
with the other nucleon being a spectator, and $\lambda=3$ to label the channel
where the two nucleons form a subsystem with the pion being a
spectator. Although we are concerned only with those physical processes having
at most one pion in either initial or final state, we do allow multiple-pion
intermediate states. These intermediate state pions may be taken to be either
distinguishable or indistinguishable without affecting the formulation below in
any essential way.

Although the derivation of the \piNN\ equations given in Ref.\ \cite{4d}
neglected all connected diagrams that are both \NN- and \piNN-irreducible in the
processes \NNNN\ and $N\!N \leftrightarrow \pi N\!N$, in our approach these are
easily included and do not complicate the original \piNN\ equations in any
essential way; for this reason we shall retain all such diagrams here. On the
other hand we follow Ref.\ \cite{4d} and keep only those \NN- and
\piNN-irreducible connected diagrams in the process \piNNpiNN\ (the three-body
forces) that are necessary to avoid undercounting. The formulation retaining
{\em all} three-body forces will be given elsewhere.

It is easy to rearrange the four-dimensional \piNN\ equations of Ref.\ \cite{4d}
into a convenient form similar to the one used by Afnan and
Blankleider~\cite{AB_80} in a three-dimensional formulation of the \piNN\
system. For the distinguishable nucleon case we obtain
\be
\T^d = \V^d + \V^d \G_t^d \T^d       \eqn{BS^d}
\ee
where $\T^d$, $\V^d$, and $\G_t^d$ are $4\times 4$ matrices
given by
\be
\T^d=\left(\begin{array}{cc} T^d_{N\!N} & \bT^d_{N} \\
T^d_{N} & T^d \end{array} \right);
\hspace{.3cm}
\V^d=\left( \begin{array}{cc} V^d_{N\!N} & \bar{\F}^d\\
\F^d & G_0^{-1}{\cal{I}} \end{array} \right);
\hspace{.3cm}
\G_t^d=\left(\begin{array}{cc} \GNN & 0 \\
0 & G_0 w^0 G_0 \end{array} \right).  \eqn{AB}
\ee
\begin{figure}[b]
\hspace*{2cm} \epsfxsize=13.5cm\epsffile{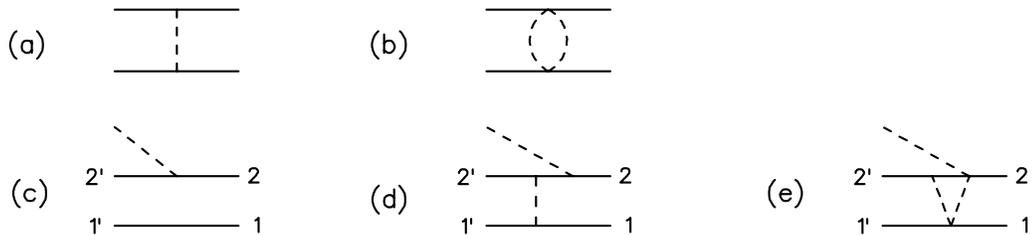}
\vspace{4mm}

\caption{\fign{v} Illustration of quantities making up $V^d_{N\!N}$.  (a)
$\VOPE{^d}$, (b) a contribution to $\VNN^{(1)}{^d}$, (c) $F_2=f_2d_1^{-1}$,
(d) $B$, and (e) a contribution to $F^d_c$.}
\end{figure}
Note that we use a notation where the superscript $d$ denotes the
distinguishable nucleon case on those symbols that will be used later without
the superscript for indistinguishable nucleons.  \eq{BS^d} is a symbolic
equation representing a Bethe-Salpeter integral equation to be solved for
$\T^d$. To clarify its meaning we give the explicit numerical form of the
equation for \NN\ scattering at the end of this subsection. $\T^d$ consists of
transition amplitudes $T^d_{N\!N}$, $T_{\lambda N}$, $T_{N \mu}$, and
$T_{\lambda\mu}$ ($\lambda$ and $\mu$ are spectator-subsystem channel labels)
the last three being elements of the matrices $T^d_N$, $\bT^d_{N}$, and
$T^d$, respectively. The physical amplitudes for \NNNN, \NNpid, \pidNN, and
\pidpid\ are then given by
\bea
\begin{array}{ccccccc}
X^d_{N\!N} = T^d_{N\!N}  & ;\hspace{2mm} &
X^d_{dN} = \bar{\psi}_dT_{3N} & ; \hspace{2mm}&
X^d_{Nd} = T_{N3}\psi_d & ; \hspace{2mm}&
X^d_{dd} = \bar{\psi}_d T_{33}\psi_d ,
\end{array}
\eea
respectively, where $\psi_d$ is the deuteron wave function in the presence of a
spectator pion.

The elements making up the kernel $\V^d$, specified in \eq{AB}, consist of
the quantities $V^d_{N\!N}$, $\F^d_\lambda$, $\bar{\F}^d_\lambda$, and
${\cal I}$ defined as follows:
\bea
V^d_{N\!N} &=& \VOPE{^d} + \VNN^{(1)}{^d}+\bar{F}_c^d G_0(F_1+F_2)
+(\bar{F}_1+\bar{F}_2)G_0F^d_c \nonumber\\
&+&\bB^dG_0B^d+\bar{F}_c^d G_0F^d_c - \Delta  \eqn{V_NN}
\eea
where $\VOPE{^d}$ is the nucleon-nucleon one pion exchange potential,
$\VNN^{(1)}{^d}$ is the \piNN-irreducible part of the \NN\ potential,
$F_i=f_id_j^{-1}$ where $i,j=1,2$ with  $i\ne j$, $f_i$ being the 
$N_i\rightarrow \pi N_i$ vertex function
and $d_j$ the Feynman propagator of nucleon $j$, $F^d_c$ is the
simultaneously \NN- and \piNN-irreducible connected amplitude for \NNtopiNN ,
$B^d=B+PBP$ where $B= \VOPE{^d} d_2 f_2$ and $P$ is the nucleon exchange
operator, and $\Delta$ is a subtraction term that eliminates
overcounting. Diagrams illustrating $\VOPE{^d}$, $\VNN^{(1)}{^d}$, $F_2$, $B$
and $F^d_c$ are given in \fig{v}. $\F^d$ is a $3\times 1$ matrix whose
$\lambda$'th row element is given by
\begin{figure}[t]
\hspace*{2cm} \epsfxsize=13.5cm\epsffile{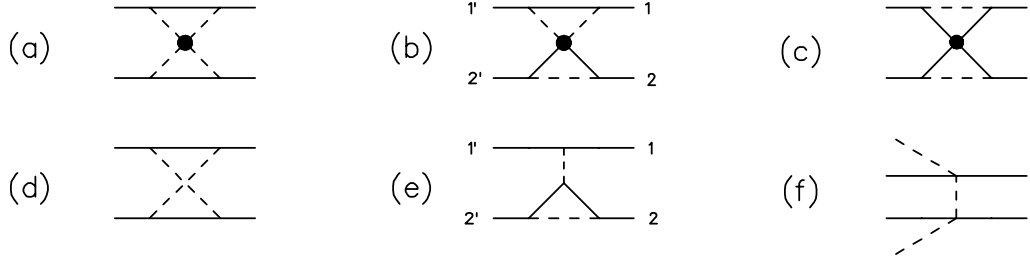}
\vspace{2mm}

\caption{\fign{x} Terms making up the subtraction term $\Delta$.  (a)
$W_{\pi\pi}$, (b) $W_{\pi N}$, (c) $W_{N\!N}$, (d) $X$, (e) $Y$, and (f) a
contribution to $F_{2\pi}$. The dark circles represent the following two-body
amplitudes: (a) full $\pi\pi$ t-matrix, (b) one-nucleon irreducible $\pi N$
t-matrix, and (c) full \NN\ t-matrix minus the \NN\ one-pion-exchange
potential.}
\end{figure}
\be
\F^d_\lambda =\sum_{i=1}^2\bar{\delta}_{\lambda i}F_i +F^d_c - B^d
\eqn{F}
\ee
\hspace*{-1mm}where $\bar{\delta}_{\lambda i} = 1-\delta_{\lambda i}$. Note that
here $B^d$ plays the role of a subtraction term.  $\bar{\F}^d$ is the $1\times
3$ matrix that is the time reversed version of $\F^d$ (similarly for other
``barred'' quantities), $G_0$ is the \piNN\ propagator, and ${\cal I}$ is the
matrix whose $(\lambda,\mu)$'th element is $\bar{\delta}_{\lambda,\mu}$.
Finally the propagator term $\G_t^d$ is a diagonal matrix consisting of the \NN\
propagator $\GNN$, and the $3\times 3$ diagonal matrix $w^0$ whose diagonal
elements are $t_1 d_2^{-1}$, $t_2 d_1^{-1}$, and $t^d_3 d_3^{-1}$, with
$t_\lambda$ being the two-body t matrix in channel $\lambda$ (for $\lambda=1$ or
$2$, $t_\lambda$ is defined to be the \piN\ $t$ matrix with the nucleon pole
term removed).
\begin{figure}[b]
\hspace*{3cm} \epsfxsize=10cm\epsffile{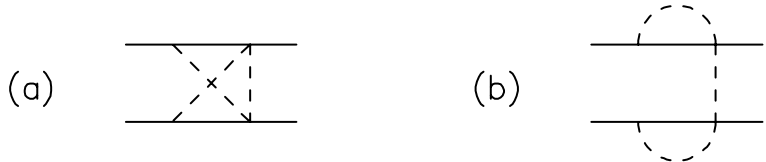}
\vspace{2mm}

\caption{\fign{f2pi} Example of diagrams contributing to
$\bar{f}_1\bar{f}_2G_0^{2\pi}F_{2\pi}$ in \protect\eq{Delta}. (a) Contribution
with crossed pions, (b) contribution with pions not crossing.}
\end{figure}
The subtraction term $\Delta$ is defined with the help of \fig{x} as follows:
\bea
\Delta &=& W_{\pi\pi} + W^d_{\pi N} + W_{N\!N} + X + Y^d \nonumber \\
&+& \bB^dG_0F^d_c
+ \bar{F}_c^d G_0 B^d + \left. \bar{F}_{2\pi} G_0^{2\pi}f_1f_2\right|\ND
+ \left. \bar{f}_1\bar{f}_2G_0^{2\pi}F_{2\pi}\right|\ND
\eqn{Delta}
\eea
where $F_{2\pi}$ is the simultaneously \NN- and \piNN-irreducible connected
amplitude for $N\!N\rightarrow \pi\pi NN$, $G_0^{2\pi}$ is the $\pi\pi NN$
propagator, $W^d_{\pi N}=W_{\pi N}+PW_{\pi N}P$ and $Y^d=Y+PYP$. The expression
$\bar{f}_1\bar{f}_2G_0^{2\pi}F_{2\pi}$ in \eq{Delta} is a subtraction term for
the overcounted contributions in $(\bF_1+\bF_2)G_0F_c^d$ in \eq{V_NN}. It
consists of two types of contributions, one with the two intermediate state
pions crossing, and one with them not crossing, as illustrated in \fig{f2pi}. As
our equations are derived with the \piNN\ vertex being dressed from the
beginning \cite{4d}, terms that contribute to a further dressing of the \piNN\
vertex, like that of \fig{f2pi}(b), need to be suppressed. This suppression is
indicated in \eq{Delta} by the ``no dressing'' symbol $|\ND$. Similar comments
of course apply to the term $\bar{F}_{2\pi} G_0^{2\pi}f_1f_2$ in \eq{Delta}.

To illustrate the numerical form of our equations we consider the case of
the \NN\ scattering amplitude $\TNN^d$ as given by \eq{BS^d} and \eq{AB}
\be
\TNN^d = \VNN^d + \VNN^d\GNN\TNN^d + \sum_{\lambda=1}^3 
\bar{\F}_\lambda^dG_0 w^0_{\lambda\lambda} G_0 T_{\lambda N}^d .
\ee
For the purposes of the illustration we may simplify this equation by neglecting
$F^d_c$ (but not $B^d$ as it is a subtraction term). In this case we obtain the
following numerical integral equation for $\TNN^d$:
\begin{figure}[t]
\hspace*{1cm} \epsfxsize=14cm\epsffile{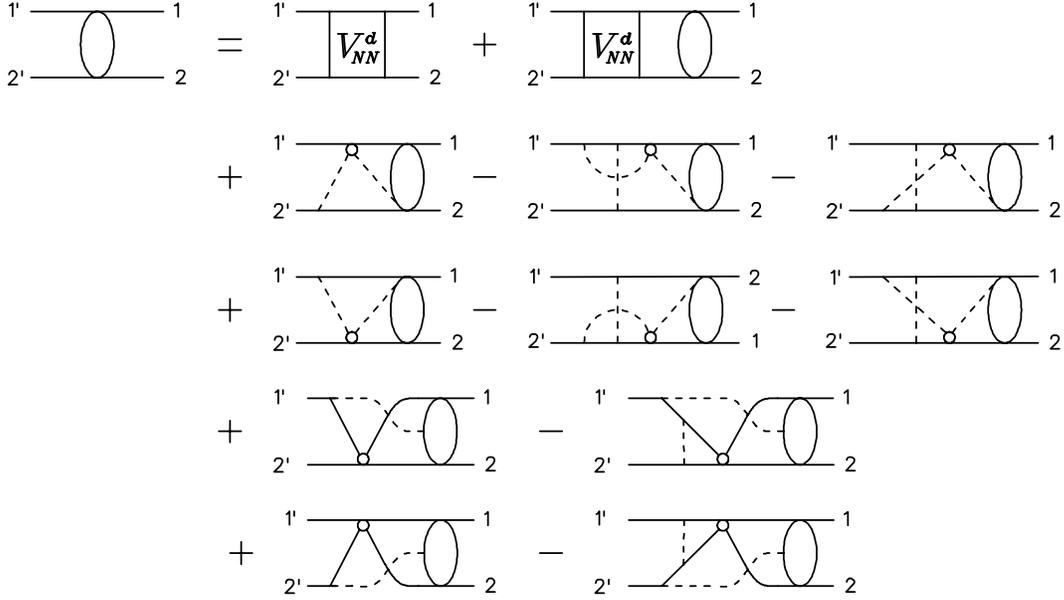}
\vspace{2mm}

\caption{\fign{tnn} Graphical representation of the integral equation, 
\protect\eq{tnn_num}, for the $N\!N$ scattering amplitude. The top line involves
the amplitude $T^d_{N\!N}$, the second line involves $T_{1N}$, the third line
involves $T_{2N}$ and the last two lines involve $T_{3N}$.}
\end{figure}
\vspace{1.1cm}

\bea
\lefteqn{\TNN^d(p_1'p_2',p_1p_2)= \VNN^d(p_1'p_2',p_1p_2)
+\int\frac{d^4k_1}{(2\pi)^4}
\VNN^d(p_1'p_2',k_1k_2)\GNN(k_1k_2)\TNN^d(k_1k_2,p_1p_2)}\nn
&&+ \int\frac{d^4k_1}{(2\pi)^4}\frac{d^4k_2}{(2\pi)^4}\left\{
\f_2(p_2',k_2k'_3)d_\pi(k'_3)t_1(p'_1k'_3,k_1k_3)\rule{0mm}{6mm}\right.\nn
&&\hspace{3cm}- \int\frac{d^4k_1'}{(2\pi)^4}
\left[\f_1(p_1',k_1''k'_3)\GpiN(k''_1k'_3)\VOPE(k_1''p_2',k_1'k_2)\right.\nn
&&\hspace{5cm}
+ \left.\f_2(p_2',k_2'k'_3)\GpiN(k'_2k'_3)\VOPE(p_1'k_2',k_1'k_2) \right]\nn 
&&\left.\hspace{5cm}\times\,d_1(k_1')t_1(k'_1k'_3,k_1k_3)\rule{0mm}{6mm}\right\}
G_0(k_1k_2k_3)
T_{1N}(k_1k_2k_3,p_1p_2) + \left(\begin{array}{c}p_1'\leftrightarrow p_2'\\
p_1\leftrightarrow p_2\end{array}\right)\nn
&&+ \int\frac{d^4k_1}{(2\pi)^4}\frac{d^4k_2}{(2\pi)^4}
\f_1(p_1',k_1'k_3)d_1(k'_1)\left[t_3(k'_1p'_2,k_1k_2)\rule{0mm}{6mm}\right.\nn
&&\hspace{4cm} \left. - \int\frac{d^4k_1''}{(2\pi)^4}
\VOPE(k_1'p_2',k_1''k_2')\GNN(k_1''k_2')t_3(k_1''k_2',k_1k_2)\right]\nn
&&\hspace{7cm}\times\, G_0(k_1k_2k_3)
T_{3N}(k_1k_2k_3,p_1p_2) + \left(\begin{array}{c}p_1'\leftrightarrow p_2'\\
p_1\leftrightarrow p_2\end{array}\right)   \eqn{tnn_num}
\eea
where to save on notation we have suppressed spin and isospin labels and used
the symbol $\left(\begin{array}{c}p_1'\leftrightarrow p_2'\\ p_1\leftrightarrow
p_2\end{array}\right)$ to indicate a term derived from the preceding
contribution by exchanging the momenta, spin, and isospin labels of the external
nucleons.  In writing \eq{tnn_num} we have used the fact that numerically
$\f_1(p',pk) =\f_2(p',pk)$, $t_1(p'k',pk)=t_2(p'k',pk)$, and
$T_{1N}(k_2k_1k_3,p_2p_1)= T_{2N}(k_1k_2k_3,p_1p_2)$. In \eq{tnn_num}
$\GpiN(k_1k_3)=d_1(k_1)d_\pi(k_3)$ is the $\pi N$ propagator, and all
unspecified momenta are determined by momentum conservation, e.g. in all the
above terms $k_3=p_1+p_2-k_1-k_2$.  \eq{tnn_num} is illustrated in \fig{tnn}. In
reference to this figure we note that despite appearances, all the subtraction
terms (terms entering with a minus sign) end on the left in the same way, either
with $\bB=\f_2d_2\VOPE$ or with $P\bB P=\f_1d_1\VOPE$. Also, the last two lines
clearly involve the contribution $t_3-\VOPE \GNN t_3$ which may appear incorrect
as the term subtracted involves an overcounted \NN\ interaction. Yet just such
an overcounted contribution needs to be subtracted to stop overcounting in the
overall equations. In this respect it is interesting to note that in the one
pion exchange ladder approximation to $t_3$, the combination $t_3-\VOPE \GNN
t_3$ reduces down simply to $\VOPE$.

\subsection{Indistinguishable nucleon case}
In approaches based on second quantisation in quantum mechanics it is usual to
obtain the scattering equations for identical particles by explicitly
symmetrizing the equations of the distinguishable particle case. This procedure
is not justified in the framework of relativistic quantum field theory.
Nevertheless, as we have already derived the \piNN\ equations taking into
account identical particle symmetry right from the beginning \cite{4d}, we can
formally deduce how the above distinguishable nucleon \piNN\ equations need to
be modified in order to get the indistinguishable nucleon case.  With this in
mind, we introduce identical nucleon transition amplitudes defined in terms of
distinguishable nucleon transition amplitudes as
\be
\begin{array}{lll}
\TNN=T^d_{N\!N}A \hspace{1.8cm} & T_{\Delta N}=T_{1N}A \hspace{.8cm} &
T_{dN}=T_{3N}A\\
T_{N\Delta }=\bar{T}_{N1}-\bar{T}_{N2}P
& T_{Nd}=\bar{T}_{N3}A
&T_{\Delta\Delta}=T_{11}-T_{12}P \\
T_{d\Delta }=T_{31}-T_{32}P &
T_{\Delta d}=T_{13}A &T_{dd}=T_{33}A
\end{array} \hspace{.2cm}   \eqn{TA}
\ee
where $A=1-P$ is the antisymmetrizing operator. As $\bar{T}_{N1}=P\bar{T}_{N2}P$
and $\bar{T}_{N3}=P\bar{T}_{N3}P$, we can alternatively write $T_{N\Delta
}=A\bar{T}_{N1}$, $T_{Nd}=A\bar{T}_{N3}$, and $T_{d\Delta}=A\bar{T}_{31}$.  Thus
in the transition to the indistinguishable particle case, the original 16
transition amplitudes for distinguishable particles have been reduced to 9
antisymmetric transition amplitudes. By taking residues of the \piNN\ Green
function for identical nucleons at two-body subsystem poles, one obtains the
following expressions for the physical amplitudes:
\bea
\begin{array}{ccccccc}
X_{N\!N} = \TNN  & ;\hspace{2mm} &
X_{dN} = \bar{\psi}_dT_{dN} & ; \hspace{2mm}&
X_{Nd} = T_{Nd}\psi_d & ; \hspace{2mm}&
X_{dd} = \bar{\psi}_d T_{dd}\psi_d .
\end{array}   \eqn{Xij}
\eea
Using \eqs{TA} in \eq{BS^d} it is easy to show that one again obtains a
Bethe-Salpeter equation
\be
\T = \V + \V\G_t\T       \eqn{BS}
\ee
but where now $\T$, $\V$, and $\G_t$ are
$3\times 3$ matrices given by
\bea
\T &=& \left( \begin{array}{ccc}
\TNN& T_{N\Delta}&T_{Nd} \\
T_{\Delta N}& T_{\Delta\Delta} & T_{\Delta d}\\ 
T_{dN}& T_{d\Delta}&T_{dd}   \end{array}  \right); \hspace{1cm}
\V = \left( \begin{array}{ccc}
V^d_{N\!N}A& A\bar{\F}^d_1 &\bar{\F}^d_3A \\
\F^d_1A & -G_0^{-1}P\hspace{2mm}  & G_0^{-1}A \\ 
\F^d_3A& G_0^{-1}A & 0   \end{array}  \right); \eqn{V} \\
\G_t &=& \left( \begin{array}{ccc}
\frac{1}{2}\GNN& 0 & 0 \\
0&G_0 t_1d_2^{-1}G_0& 0\\ 
0 & 0 & \frac{1}{4}G_0t_3d_3^{-1}G_0   \end{array}  \right) \eqn{G}
\eea
where $t_3=t^d_3A$ is the $t$ matrix for two identical nucleons in the presence
of a spectator pion.

From \eq{V_NN} it follows that
\bea
V^d_{N\!N}A&=&\VOPE + \VNN^{(1)} + \bar{F}_c G_0F_1^{R}
+ \bar{F}_1^{L}G_0 F_c \nonumber \\
&+& \bar{F}_{2\pi}^L G_0^{2\pi}f_1f_2 + \bar{f}_1\bar{f}_2G_0^{2\pi}F_{2\pi}^R
- W_{\pi\pi}^R - W_{\pi N}^{LR}- W_{N\!N}^R-X^R - Y^{LR}
\nonumber\\
&+&\frac{1}{2}(\bB^{LR}-\bar{F}_c )G_0(B^{LR}-F_c)  \eqn{V_NN_A}
\eea 
where $\VOPE=\VOPE{^d}A$, $\VNN^{(1)}=\VNN^{(1)d}A$, $F_c=F^d_c A$,
$\bar{F}_c=\bar{F}_c^d A$, and where we have used a superscript $L$ to
indicate that $A$ is acting on the left, and a superscript $R$ to indicate that
$A$ is acting on the right. Note that $W_{\pi\pi}^R=W_{\pi\pi}^L$, 
$W_{N\!N}^R=W_{N\!N}^L$,  $X^R=X^L$, and  $F_{2\pi}^R=F_{2\pi}^L$.
Here we have also used the simple result
\be
(\bB^d-\bar{F}_c^d)G_0(B^d-F^d_c)A
=\frac{1}{2}(\bB^{LR}-\bar{F}_c)G_0(B^{LR}-F_c).
\ee
The other terms in the kernel $\V$ of \eq{V} are
\bea
\F^d_1 &=& F_2-B^d+F^d_c;\hspace{1.9cm}
\bar{\F}^d_1=\bar{F}_2-\bB^d+\bar{F}_c^d,\\
\F^d_3&=& F_1+F_2-B^d+F^d_c;
\hspace{1cm}\bar{\F}^d_3=\bF_1+\bF_2-\bB^d+\bar{F}_c^d.
\eqn{VWk}
\eea

The \piNN\ equations as given by \eqs{BS}, (\ref{V}) and (\ref{G}) have a form
where seven of the eight non-zero elements making up the kernel $\V$ contain the
antisymmetrization operator $A$. One element ($t_3$) in $\G_t$ also contains
$A$. The question arises if one can remove the operators $A$ from these
equations by analogy with the familiar example of the one-channel Bethe-Salpeter
equation for two identical nucleons:
\be
T=VA(1 + \frac{1}{2}GT).      \eqn{BS-1chan}
\ee
Here the \NN\ potential $V$ is multiplied on the right by $A$, but as $V$ has
the property $VP=PV$ and therefore $VA=AV$, it is easy to see that we can
instead solve the equation
\be
T'=V(1+GT')
\ee
where $A$ has been removed from the equation and where the \NN\ $t$ matrix $T$
is obtained by antisymmetrizing the solution: $T=T'A$. 

Unfortunately it is not possible to follow a similar procedure to completely
remove the operator $A$ from \eqs{BS}, (\ref{V}) and (\ref{G}). The reason is
that $\V P\ne P\V$ for the kernel $\V$ of \eq{V} (in particular $\F^d_1P\ne
P\F^d_1$). In this respect we note that the \piNN\ equations for identical
particles given in Ref. \cite{PA4d} are not equivalent to ours as they do not
involve the operator $A$. Although $A$ cannot be removed completely from the
identical particle \piNN\ equations, one can reduce the number of places where
$A$ appears. For example, it is easy to show that \eqs{BS}, (\ref{V}) and
(\ref{G}) are equivalent to the equations
\be
\T' = \V' + \V'\G_t'\T' ;
\ee
\be
\V' = \left( \begin{array}{ccc}
V^d_{N\!N}& \bar{\F}^d_1 &\bar{\F}^d_3 \\
\F^d_1 A & -G_0^{-1}P\hspace{2mm}  & G_0^{-1} A\\ 
\F^d_3 & G_0^{-1} & 0   \end{array}  \right); \hspace{3mm}
\G_t' = \left( \begin{array}{ccc}
\GNN& 0 & 0 \\
0&G_0 t_1d_2^{-1}G_0& 0\\ 
0 & 0 & G_0t_3^d d_3^{-1} G_0   \end{array}  \right)
\ee
where
\be
\T=\left(\begin{array}{ccc}  A & 0 & 0\\
                             0 & 1 & 0\\
                             0 & 0 & A \end{array}\right) T'
\left(\begin{array}{ccc} \Ah & 0 & 0\\
                           0 & 1 & 0\\
                           0 & 0 & \Ah\end{array}\right)
\ee
and all but two operators $A$ have been removed form the integral equations.
There are further forms of the \piNN\ equations with operators $A$ appearing
at various other places in the equations, yet there are always at least
two $A$ operators present.

\section{\mbox{\boldmath{$\pi N\!N$}} Electromagnetic transition currents}

\subsection{Gauging the \mbox{\boldmath{$\pi N\!N$}} equations}

In this section we shall derive expressions for the various electromagnetic
transition currents of the \piNN\ system. To do this we utilise the recently
introduced gauging of equations method \cite{gnnn4d,g4d}. As the gauging
procedure is identical for the distinguishable and indistinguishable particle
cases, we restrict our attention to the \piNN\ system where the nucleons are
treated as indistinguishable particles.

As discussed in the previous section, the strong interaction \piNN\ equations
can be written in a number of equivalent forms. Choosing the form given by
\eqs{BS}, (\ref{V}) and (\ref{G}), direct gauging of \eq{BS} gives
\be
\T^\mu = \V^\mu + \V^\mu \G_t\T + \V\G_t^\mu\T + \V\G_t\T^\mu
\ee
which can easily be solved for $\T^\mu$ giving
\be
\T^\mu = (1+\T\G_t)\V^\mu(1+\G_t\T) + \T \G_t^\mu \T \eqn{T^mu}.
\ee
$\T^\mu$ is a matrix of gauged transition amplitudes $T^\mu_{N\!N}$,
$T^\mu_{N\Delta}$, $T^\mu_{Nd}$, etc. To obtain the physical electromagnetic
transition currents of the \piNN\ system where photons are attached everywhere
it is not sufficient to just gauge the physical \piNN\ amplitudes of \eq{Xij}.
Although this would indeed attach photons everywhere inside the strong
interaction diagrams, it would miss those contributions to the physical
electromagnetic transition currents where the photons are attached to the
external (initial and final state) pions and nucleons.  In order to also include
these external leg contributions it is useful to attach the corresponding
propagators to the $X$-amplitudes of \eq{Xij}:
\bea
\tX_{N\!N} &=&  \GNN X_{N\!N} \GNN \hspace{1cm}
\tX_{dN} =  d_\pi X_{dN} \GNN    \eqn{tX1} \\
\tX_{Nd} &=& \GNN X_{Nd} \, d_\pi  \hspace{1.5cm}
\tX_{dd} = d_\pi X_{dd}\, d_\pi  .  \eqn{tX2}
\eea
The physical electromagnetic transition currents are then obtained by gauging
\eqs{tX1} and (\ref{tX2}) and ``chopping off'' external legs:
\bea
j^\mu_{N\!N} &=& \GNN^{-1}\tX_{N\!N}^\mu \GNN^{-1}\hspace{1cm}
j^\mu_{dN} = d_\pi^{-1} \tX_{dN}^\mu \GNN^{-1} \eqn{X^mu1}\\
j^\mu_{Nd} &=& \GNN^{-1}\tX_{Nd}^\mu\, d_\pi^{-1}\hspace{1.3cm}
j^\mu_{dd} = d_\pi^{-1} \tX_{dd}^\mu \,d_\pi^{-1}.  \eqn{X^mu2}
\eea
Using \eqs{Xij} we obtain that
\bea
j^\mu_{N\!N} &=& \GNN^{-1}D_0^\mu\TNN
+ \TNN D_0^\mu\GNN^{-1} + \TNN^\mu \eqn{j_NN^mu}\\
j^\mu_{dN} &=& \bphi_d^\mu \GNN T_{dN} + d_\pi^{-1}\bphi_d G_0^\mu T_{dN}
+ \bphi_d \GNN T^\mu_{dN} \nonumber \\
&+& \bphi_d \GNN T_{dN} \GNN^\mu \GNN^{-1}\eqn{j_dN^mu}\\
j^\mu_{Nd} &=& T_{Nd}\GNN \phi_d^\mu  + T_{Nd}G_0^\mu \phi_d d_\pi^{-1}
+  T^\mu_{Nd}  \GNN\phi_d \nonumber \\
&+& \GNN^{-1} \GNN^\mu T_{Nd} \GNN \phi_d  \eqn{j_Nd^mu} \\
j^\mu_{dd} &=& \bphi_d^\mu \GNN T_{dd}\GNN\phi_d
+ d_\pi^{-1}\bphi_d \GNN^\mu T_{dd}\GNN\phi_d \nonumber \\
&+& \bphi_d \GNN T_{dd}^\mu \GNN\phi_d
+ \bphi_d \GNN T_{dd}\GNN^\mu\phi_d d_\pi^{-1}\nonumber \\
&+& \bphi_d \GNN T_{dd}\GNN\phi_d^\mu \eqn{j_dd^mu}
\eea
where $\phi_d$ is the deuteron bound state vertex function defined by the
relation $\psi_d=d_1d_2\phi_d$. The above equations express the physical
electromagnetic transition currents $j^\mu_{\A\B}$ ($\A,\B = N\hspace{2mm}
\mbox{or}\hspace{2mm} d$) in terms of the half-on-shell \piNN\ transition
amplitudes $T_{\A\B}$ and the gauged quantities $V^\mu_{\A\B}$,
$(\G^\mu_t)_{\A\B}$, and $\phi_d^\mu$. Note that $\phi_d^\mu$ consists of
contributions where the photon is attached everywhere inside the deuteron bound
state, and is determined by gauging the two-nucleon bound state equation for
$\phi_d$ \cite{gnnn4d,g4d}.

That the above equations are gauge invariant is evident from the fact that we
have formally attached the photon to all possible places in the strong
interaction model. The gauge invariance of our equations also follows from a
strict mathematical proof; however, as this proof is essentially identical to
the one given for the \NNN\ system \cite{g4d}, we shall not repeat it here.

\subsection{Alternative form of the \mbox{\boldmath{$\pi N\!N$}} equations}

Although the preceding discussion solves the problem of gauging the \piNN\
system, the expression obtained for the gauged transition amplitudes, \eq{T^mu},
may not be the most convenient for numerical calculations. The disadvantage of
\eq{T^mu} is that it utilises a Green function $\G_t$ which contains two-body
$t$ matrices, while such $t$ matrices are already implicitly present in the
adjoining amplitudes of $\T$. This makes the calculation of $\T^\mu$
unnecessarily complicated. One could eliminate this multiple $t$-dependence in
\eq{T^mu} by making use of \eq{BS}; however, this would be a lengthy and awkward
procedure.  Instead we derive an alternative form of the \piNN\ equations which
uses a ``free'' Green function which contains no two-body interactions and which
leads to simpler expressions for the \piNN\ electromagnetic transition currents.

The \piNN\ equations \eqs{BS}, (\ref{V}) and (\ref{G}), can be written in the
form
\be \left(
\begin{array}{cc} \TNN & \bT_{N} \\
T_{N} & T \end{array} \right)= \left( \begin{array}{cc}
\VNN & \bar{\F} \\
\F & L G_0^{-1} \end{array} \right)\left[ 1+ \left(
\begin{array}{cc} \frac{1}{2}\GNN & 0 \\
0 & G_0tG_0 \end{array} \right) \left(
\begin{array}{cc} \TNN & \bT_{N} \\
T_{N} & T \end{array} \right) \right].  \eqn{KB}
\ee
where $\VNN=V^d_{N\!N}A$ is given by \eq{V_NN_A}, and
\bea
\F &=&\left( \begin{array}{c}
\F^d_1A \\ \F^d_3A \end{array}  \right)
=\left[ \begin{array}{c} (F_2-B^d+F^d_c)A \\
(F_1+F_2-B^d+F^d_c)A \end{array}  \right]
=\left[ \begin{array}{c} F_2A-B^{LR}+F_c \\
(F_1+F_2)A-B^{LR}+F_c \end{array}  \right] , \\
\nonumber\\
\bar{\F} &=&\left(A\bar{\F}^d_1\hspace{2mm} A\bar{\F}^d_3\right)
= \left[A(\bar{F}_2-\bB^d+\bar{F}_c^d)\hspace{3mm} 
A(\bF_1+\bF_2-\bB^d+\bF_c^d)\right] \nonumber \\
&=& \left[(A\bar{F}_2-\bB^{LR}+\bar{F}_c)
\hspace{3mm} (A\bar{F}_1+A\bar{F}_2-\bB^{LR}+\bar{F}_c)\right],  \\
\nonumber\\
L&=&\left( \begin{array}{cc}
-P  & A \\ 
A & 0   \end{array}  \right), \hspace{.7cm}t=\left( \begin{array}{cc}
t_1d_2^{-1}  &  0 \\ 
0    & \frac{1}{4}t_3d_3^{-1}  \end{array}  \right).  \eqn{gkb3}
\eea
With the view of gauging Green function versions of \piNN\ transition
amplitudes, we introduce the Green function matrix appropriate to \eq{KB}:
\bea
\G &=& \left(\begin{array}{cc} G_{N\!N} & \bar{G}_{N} \\
G_{N} & G \end{array} \right)
\equiv
\left( \begin{array}{ccc}
G_{N\!N}& G_{N\Delta}&G_{Nd} \\
G_{\Delta N}& G_{\Delta\Delta} & G_{\Delta d}\\ 
G_{dN}& G_{d\Delta}&G_{dd}  \end{array}  \right) \nonumber \\
&=&\left(\begin{array}{cc} A\GNN & 0 \\
0 & 0 \end{array} \right)+\left(
\begin{array}{cc} \GNN & 0 \\
0 & G_0 \end{array} \right)
\left(\begin{array}{cc} \TNN & \bT_{N} \\ 
T_{N} & T \end{array} \right)
\left(\begin{array}{cc} \GNN & 0 \\
0 & G_0 \end{array} \right) . \eqn{G1}
\eea
The inhomogeneous term is chosen so that $G_{N\!N}$ corresponds
exactly to the full Green function for \NN\ scattering.  Then, as shown in the
Appendix, $\G$ satisfies the equation
\be
\G=\left( \begin{array}{cc}
A\GNN & 0 \\
0  & L G_0 \end{array} \right)
\left[1+\left( \begin{array}{cc}
\frac{1}{4}\VNN-
\frac{1}{4}\bar{\Lambda}L G_0\Lambda& 
\hspace{2mm}\frac{1}{2}\bar{\Lambda} \\
\frac{1}{2}\Lambda & \hspace{2mm}t \end{array} \right)\G\right]
\eqn{G2}
\ee
where $\Lambda$ and $\bar{\Lambda}$ are defined by \eqs{Lam}.  The essential
feature of \eq{G2} is that it is written in terms of an effective ``free'' Green
function matrix
\be
\G_0 = \left( \begin{array}{cc}
A\GNN & 0 \\
0  & L G_0 \end{array} \right)
\ee
which does not involve two-body interactions. For this reason \eq{G2} is ideal
for the purposes of gauging.

\subsection{Gauging the alternative form of the \mbox{\boldmath{$\pi N\!N$}}
equations}

In terms of the elements of $\G$, the Green function versions of the
physical amplitudes [defined in \eq{tX2}] are given by
\bea
\begin{array}{ccccccc}
\tX_{N\!N} = G_{N\!N}  & ;\hspace{2mm} &
\tX_{dN} = \bar{\phi}_dG_{dN} & ; \hspace{2mm}&
\tX_{Nd} = G_{Nd}\phi_d & ; \hspace{2mm}&
\tX_{dd} = \bar{\phi}_d G_{dd}\phi_d .
\end{array}
\eea
After gauging, these equations give
\bea
\tX_{N\!N}^\mu &=& G_{N\!N}^\mu \eqn{tX^mu1} \\
\tX_{dN}^\mu &=& \bar{\phi}_d^\mu G_{dN}+\bar{\phi}_dG_{dN}^\mu \\
\tX_{Nd}^\mu &=& G_{Nd}^\mu\phi_d+G_{Nd}\phi_d^\mu \\
\tX_{dd}^\mu &=& \bar{\phi}_d^\mu G_{dd}\phi_d+\bar{\phi}_d G_{dd}^\mu\phi_d
+\bar{\phi}_d G_{dd}\phi_d^\mu. \eqn{tX^mu4}
\eea
The \piNN\ electromagnetic transition currents $j^\mu_{\A\B}$ are then
determined by \eqs{X^mu1} and \eqs{X^mu2}.

To determine the quantities $G_{\A\B}^\mu$ in \eqs{tX^mu1}-(\ref{tX^mu4})
we need to derive the expression for $\G^\mu$ by gauging \eq{G2}. Defining
\be
\V_t=\left( \begin{array}{cc}
\frac{1}{4}\VNN-
\frac{1}{4}\bar{\Lambda}L G_0\Lambda& 
\hspace{2mm}\frac{1}{2}\bar{\Lambda} \\
\frac{1}{2}\Lambda & \hspace{2mm}t \end{array} \right)
\ee
\eq{G2} can be written as
\be
\G = \G_0+\G_0 \V_t \G.   \eqn{G3}
\ee
Gauging this equation and solving for $\G^\mu$ gives
\be
\G^\mu=(1+\G\V_t)\G_0^\mu(1+\V_t\G)+\G\V_t^\mu\G.   \eqn{G^mu_tmp}
\ee
To simplify this equation we cannot use \eq{G3} to write
$1+\V_t\G=\G_0^{-1}\G$ and $1+\G\V_t=\G\G_0^{-1}$ since $L$ is
singular so that the inverse $\G_0^{-1}$ does not exist. Instead we use the
fact that
$L=L\Omega L$, where
\be
\Omega= \left( \begin{array}{cc}
-\frac{1+P}{2}  & \frac{1}{2} \\ 
\frac{1}{2} & -\frac{1}{4}   \end{array}  \right)
\ee
which allows us to write
\be
\G_0^\mu = \G_0 \M^\mu \G_0  \eqn{G_0^mu}
\ee
where
\be
\M^\mu=\left( \begin{array}{cc}
\frac{1}{2}\GNN^{-1}\GNN^\mu \GNN^{-1} & 0 \\
0  & \Omega G_0^{-1}G_0^\mu G_0^{-1} \end{array} \right).
\ee
Using \eq{G_0^mu} in \eq{G^mu_tmp} allows us to write a compact expression for
$\G^\mu$:
\be
\G^\mu = \G\left( \M^\mu + \V_t^\mu \right) \G.    \eqn{G^mu}
\ee
Comparing this expression with the one of \eq{T^mu}, we see that both involve
the gauged two-body $t$ matrix $t^\mu$; however, in contrast to \eq{T^mu}, the
above expression does not contain adjoining strong interaction $t$ matrices and
may therefore be preferable for numerical calculations.

As $\GNN^\mu=(d_1d_2)^\mu = d_1^\mu d_2 +d_1 d_2^\mu$ and
$G_0^\mu=(d_1d_2d_3)^\mu = d_1^\mu d_2 d_3 + d_1 d_2^\mu d_3 +
d_1 d_2 d_3^\mu$, we obtain
\bea
\GNN^{-1}\GNN^\mu\GNN^{-1} &=& \Gamma_1^\mu d_2^{-1}+d_1^{-1}\Gamma_2^\mu\\
G_0^{-1}G_0^\mu G_0^{-1} &=& \Gamma_1^\mu d_2^{-1}d_3^{-1}
+d_1^{-1}\Gamma_2^\mu d_3^{-1}+d_1^{-1}d_2^{-1}\Gamma_3^\mu
\eea
where
\be
\Gamma_i^\mu = d_i^{-1} d_i^\mu d_i^{-1}
\ee
is the electromagnetic vertex function of particle $i$. The term $\M^\mu$
of \eq{G^mu} thus corresponds to photon coupling in the impulse approximation.
The gauged matrix $\V_t^\mu$ corresponds to the interaction currents and
consists of the elements $V^\mu=(\VNN-\bar{\Lambda}LG_0\Lambda)^\mu$,
$\bar{\Lambda}^\mu$, $\Lambda^\mu$ and $t^\mu$. It is important to note that the
diagonal elements of matrix $t^\mu$ are both of the form
\be
\left(t_id_j^{-1}\right)^\mu = t_i^\mu d_j^{-1}+ t_i \left(d_j^{-1}\right)^\mu
=t_i^\mu d_j^{-1} - t_i\Gamma_j^\mu
\ee
where the last equality follows from the fact that
$\left(d_j^{-1}d_j\right)^\mu=0$. Thus the diagonal elements of $t^\mu$ involve
new subtraction terms $t_i\Gamma_j^\mu$ whose origin does not lie in the
subtraction terms of the strong interaction \piNN\ equations, but rather in
the gauging procedure itself. Analogous subtraction terms arise
in the three-nucleon problem whose strong interaction equations have no
subtraction terms \cite{gnnn4d,g4d}. Similar subtraction terms will arise in the
gauging of $F_1$ and $\bF_1$ contained in $\Lambda$ and $\bar{\Lambda}$
respectively. The graphical representation of $\M^\mu$ and $\V_t^\mu$ is
given in \fig{gmu}.
\begin{figure}[t]
\epsfxsize=17.0cm\epsffile{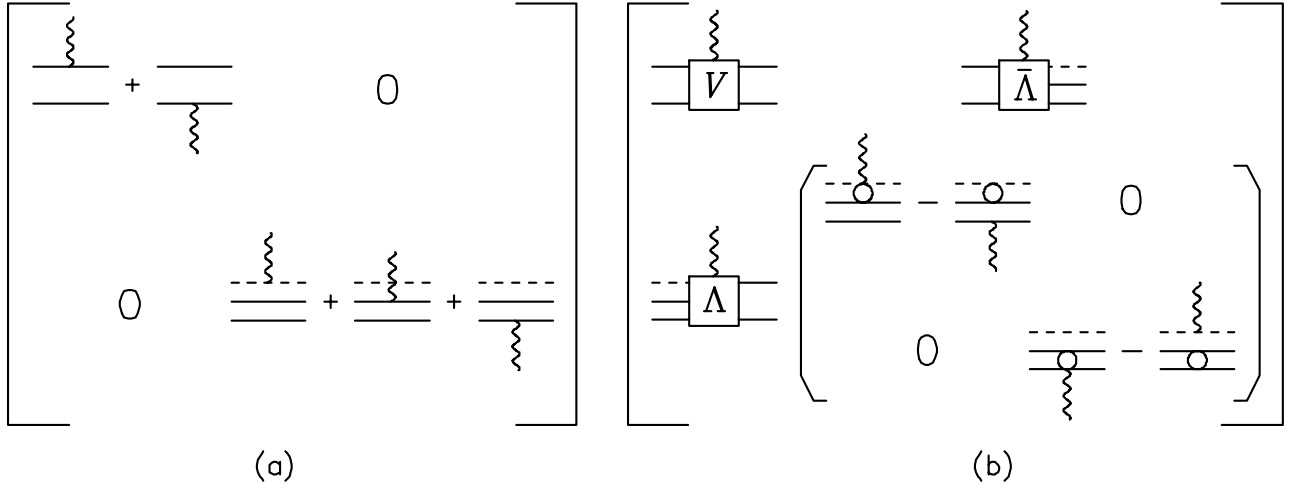}
\vspace{2mm}

\caption{\fign{gmu} Graphical representation of (a) $\M^\mu$ and (b) $\V_t^\mu$
which enter into the expression for the gauged Green function matrix $\G^\mu$ of
\protect\eq{G^mu}. Constant factors and effects of operator $A$ and matrix
$\Omega$ have been ignored in this illustration.}
\end{figure}

\eq{G^mu} can be used to determine all the possible electromagnetic transition
current of the \piNN\ system. For example,
current $j^\mu_{N\!N}$ for the electromagnetic transition \NNNN\ is given by
\bea
j^\mu_{N\!N} &=& \GNN^{-1}G_{N\!N}^\mu\GNN^{-1}\\
&=&\sum_{\A\B} \GNN^{-1}G_{N\A}\left(\M^\mu+\V_t^\mu\right)_{\A\B}
G_{\B N}\GNN^{-1}\nn
&=&\sum_{\A\B} \left(A\delta_{N\A}+T_{N\A}D_\A\right)
\left(\M^\mu+\V_t^\mu\right)_{\A\B}
\left(A\delta_{\B N}+D_\B T_{\B N}\right)
\eea
where $\A,\B=N,\Delta,d$, and $D_N\equiv \GNN$, $D_\Delta=D_d\equiv G_0$.

An especially interesting use of \eq{G^mu} is to study the electromagnetic
properties of bound states of the \piNN\ system. It is certainly expected
that the strong interaction \piNN\ model under discussion admits a bound state
corresponding to the physical deuteron. In this case a solution will exist to
the homogeneous version of \eq{BS}:
\be
\Phi = \V\G_t\Phi   \eqn{Phi}
\ee
where $\Phi$ is a matrix of deuteron vertex functions
\be
\Phi = \left(\begin{array}{c} \Phi_N \\ \Phi_\Delta \\
\Phi_d \end{array}\right).
\ee
Here $\Phi_N$ is the usual deuteron vertex function describing the $d\rightarrow
N N$ transition, while $\Phi_\Delta$ and $\Phi_d$ are somewhat unusual in that
they describe transitions to clustered \piNN\ states: $d\rightarrow (\pi N) N$
and $d\rightarrow (N N) \pi$ respectively. Comparing \eq{G1} and \eq{G2} it is
seen that $\G$ has a pole at the deuteron mass $M_d$:
\be
\G \sim i \frac{\Psi\bar{\Psi}}{P^2-M_d^2}\hspace{1cm}\mbox{as}\hspace{.5cm}
P^2\rightarrow M_d^2    \eqn{G_pole} 
\ee
where $P$ is the total four-momentum of the system,
and where $\Psi$ satisfies the equation
\be
\Psi=\G_0\V_t\Psi.   \eqn{Psi}
\ee
Clearly $\Psi$ is related to $\Phi$ by the equation
\be
\Psi = \left(\begin{array}{cc} \GNN & 0 \\
0 & G_0 \end{array}\right) \Phi
\ee
so that either of the equations (\ref{Phi}) or (\ref{Psi}) can be used to
determine $\Psi$.
Taking the left and right residues at the deuteron bound state poles of
\eq{G^mu} we obtain the bound state electromagnetic current
\be
j^\mu = \bar{\Psi}\left(\M^\mu+\V_t^\mu\right)\Psi     \eqn{j^mu}
\ee
which describes the electromagnetic properties of the deuteron whose internal
structure is described by the present \piNN\ model. \eq{j^mu} provides a very
rich description of the internal electromagnetic structure of the deuteron with
all possible meson exchange currents being taken into account in a gauge
invariant way. In view of the accuracy of this model which is based on meson and
baryon degrees of freedom, a comparison of the deuteron electromagnetic form
factors (easily extracted from $j^\mu$) with those extracted from experiment
should prove to be most interesting.

\section{Conclusions}

We have derived gauge invariant expressions for the electromagnetic transition
currents of the \piNN\ system where the strong interactions are described by
four-dimensional integral equations. The feature of our approach is that the
external photon is coupled everywhere in the strong interaction model, in this
way giving a unified description of the \piNN\ and $\gamma$\piNN\ systems.  This
unified approach to the \piNN\ system and its electromagnetic currents has been
made possible by the recent introduction of the gauging of equations method
\cite{gnnn4d,g4d}. The use of this method has also enabled us to avoid all the
overcounting problems that are inherent in four-dimensional descriptions of
\piNN -like systems \cite{4d}.

The expressions we have derived can be used directly to make four-dimensional
calculations of all the reactions induced in the \piNN\ system by an external
electromagnetic probe. However, in view of the practical difficulty of solving
four-dimensional integral equations, it may also be useful to have a gauge
invariant three-dimensional description of the same processes. In that case our
four-dimensional expressions can be used to provide the starting point for a
three-dimensional reduction. One such three-dimensional reduction scheme
that preserves gauge invariance and is easily applied to our four-dimensional
expressions was discussed in Ref.\ \cite{g3d}.

Although we have specifically gauged the \piNN\ system, it should be noted that
the derived expressions apply equally well to other systems consisting of two
fermions and one boson which can be absorbed by the fermions. For example one
could apply our expressions to the quark-antiquark-gluon system in order to
calculate the meson spectrum including its electromagnetic properties.

It is also worth noting that our derivation of the gauged \piNN\ equations only
assumed that the external field couples to hadrons only to first order in the
field-hadron coupling constant, but otherwise does not depend on the nature of
the external field. Thus, for example, our expressions can be used directly to
determine the weak interaction transition currents of the \piNN\ system.

\section*{Acknowledgements}
The authors would like to thank the Australian Research Council for their
financial support.

\section*{Appendix}

In this Appendix we derive \eq{G2}.
Writing \eq{KB} as $\T=\V+\V\G_t\T$ and \eq{G1} as $\G=\G_u+g\T g$ where
\be
\G_u=\left(\begin{array}{cc} A\GNN & 0 \\
0 & 0 \end{array} \right),\hspace{10mm}
g=\left(
\begin{array}{cc} \GNN & 0 \\
0 & G_0 \end{array} \right),     \eqn{ap_1}
\ee
\be
\V=\left( \begin{array}{cc}
\VNN & \bar{\F} \\
\F & L G_0^{-1} \end{array} \right),\hspace{10mm}
\G_t=\left(
\begin{array}{cc} \frac{1}{2}\GNN & 0 \\
0 & G_0tG_0 \end{array} \right) ,    \eqn{ap_2}
\ee
we have that
\be
\G=\G_u+g\T g=\G_u+g\V g+g\V\G_t\T g=\G_u+g\V g+g\V\G_tg^{-1}(\G-\G_u)
\ee
or
\be
\G=\G_u+g\V \left(g-\G_t g^{-1}\G_u\right)+g\V\G_t g^{-1}\G.
\ee
Using the explicit matrix forms for $\G_u$, $g$, $\V$, and $\G_t$ given in
\eq{ap_1} and \eq{ap_2}, the equation for $\G$ takes the form
\be
\G=\left(\begin{array}{cc} A\GNN\hspace{1mm} &\GNN \bar{\F} G_0 \\
0 & L G_0 \end{array} \right)+\left( \begin{array}{cc}
\frac{1}{2}\GNN \VNN\hspace{1mm}  & \GNN\bar{\F} G_0t \\
\frac{1}{2}G_0\F & L G_0t \end{array} \right)\G .
\ee
Multiplying this equation from the left by the matrix 
\be
R=\left(\begin{array}{cc} 1 &\hspace{2mm}
-\GNN\bar{\F} Z \\
0 & ZL \end{array} \right)
\ee
where 
\be
Z=\left( \begin{array}{cc}
-\frac{1+P}{2}  & \frac{A}{4} \\ 
\frac{A}{4} & -\frac{A}{8}   \end{array}  \right)\hspace{10mm}\mbox{so that}
\hspace{1cm}
ZL=L Z=\left( \begin{array}{cc}
1  & 0 \\ 
0  & \frac{A}{2}   \end{array}  \right),
\ee
we obtain
\be
\left(\begin{array}{cc} 1 & \hspace{2mm}-\GNN \bar{\F} Z \\
0 & ZL \end{array} \right)\G =
\left( \begin{array}{ccc} A\GNN& 0  \\
0  & L G_0   \end{array}  \right) +
\left( \begin{array}{ccc} \GNN& 0  \\
0  & L G_0   \end{array}  \right)
\left( \begin{array}{cc}
\frac{1}{2}\VNN-\frac{1}{2}\bar{\F} G_0Z\F\hspace{2mm} &  0 \\
\frac{1}{2}Z\F & t \end{array} \right)\G  .  \eqn{ap_3}
\ee
Transferring the off-diagonal term $-D_0\bar{\F} Z$ from the left
hand side to the right and recognising that
\be
\left(\begin{array}{cc} 1 & 0 \\
0 \hspace{1mm}& L Z \end{array} \right) \G = \G
\ee
\eq{ap_3} becomes
\be
\G=\left( \begin{array}{cc} A\GNN&
 0 \\
0 & L G_0 \end{array} \right)
 +\left( \begin{array}{cc} \GNN&
 0 \\
0 & L G_0 \end{array} \right)\left( \begin{array}{cc}
\frac{1}{2}\VNN-\frac{1}{2}\bar{\F} G_0Z\F\hspace{1mm} & 
 \bar{\F} Z \\
\frac{1}{2}Z\F & t \end{array} \right) \G .   \eqn{ap_4}
\ee
This equation may be written in the form
\be
\G=\left( \begin{array}{cc}
A\GNN & 0 \\
0  & L G_0 \end{array} \right)
\left[1+\left( \begin{array}{cc}
\frac{1}{4}\VNN-
\frac{1}{4}\bar{\Lambda}L G_0\Lambda& 
\hspace{2mm}\frac{1}{2}\bar{\Lambda} \\
\frac{1}{2}\Lambda & \hspace{2mm}t \end{array} \right)\G\right]. \eqn{ap_5}
\ee
where
\be
\Lambda = \left\{\begin{array}{c}
\left[F_1-{\textstyle\frac{1}{2}}(B^d-F_c^d)\right]A \\
{\textstyle -\frac{1}{2}}(B^d-F_c^d)
\end{array} \right\} ; \hspace{1cm}
\bar{\Lambda} =
\left\{A\left[\bF_1-{\textstyle\frac{1}{2}}(\bB^d-\bF_c^d)\right]
\hspace{3mm}
{\textstyle -\frac{1}{2}}(\bB^d-\bF_c^d)\right\}. \eqn{Lam}
\ee
We note that the presence of the antisymmetrization operator $A$ in the term
$AD_0$ of \eq{ap_5} allows one to eliminate the antisymmetrization operators in
$\VNN$ and $\bar{\Lambda}$ by making the replacements
\be
\VNN\rightarrow 2\VNN^d\hspace{1cm}\mbox{and}\hspace{1cm}
A\left[\bF_1-{\textstyle\frac{1}{2}}(\bB^d-\bF_c^d)\right]\rightarrow
2\left[\bF_1-{\textstyle\frac{1}{2}}(\bB^d-\bF_c^d)\right]
\ee
in \eq{ap_5} and \eq{Lam} respectively.

\end{document}